\documentclass[12pt, onecolumn, draftcls]{IEEEtran}
\usepackage{times}
\usepackage{epsf,epsfig}
\usepackage{cite}
\usepackage{amsmath, alltt, amssymb, xspace}

\begin{document}

\title{Power-Efficient Direct-Voting  Assurance for Data Fusion in Wireless Sensor Networks
\thanks{The preliminary result of this work is presented at the
IEEE International Conference on Sensor Networks, Ubiquitous, and
Trustworthy Computing (SUTC2006), Taiwan, R.O.C.. This work was
supported by the National Science Council of Taiwan, R.O.C., under
grants NSC 94-2213-E-305-002 and NSC 94-2213-E-305-001.}}

\author{Hung-Ta~Pai and Yunghsiang~S.~Han \thanks{The authors are
with the Graduate Institute of Communication Engineering, National
Taipei University, Taiwan (E-mail: htpai@mail.ntpu.edu.tw;
yshan@mail.ntpu.edu.tw).}}

\maketitle \thispagestyle{empty}

\setlength{\baselineskip}{24pt}
\begin{abstract}
\noindent Wireless sensor networks place sensors into an area to
collect data and send them back to a base station. Data fusion,
which fuses the collected data before they are sent to the base
station, is usually implemented over the network. Since the sensor
is typically placed in locations accessible to malicious attackers,
information assurance of the data fusion process is very important.
A witness-based approach has been proposed to validate the fusion
data. In this approach, the base station receives the fusion data
and "votes" on the data from a randomly chosen sensor node. The vote
comes from other sensor nodes, called "witnesses," to verify the
correctness of the fusion data. Because the base station obtains the
vote through the chosen node, the chosen node could forge the vote
if it is compromised. Thus, the witness node must encrypt the vote
to prevent this forgery. Compared with the vote, the encryption
requires more bits, increasing transmission burden from the chosen
node to the base station. The chosen node consumes more power. This
work improves the witness-based approach using direct voting
mechanism such that the proposed scheme has better performance in
terms of assurance, overhead, and delay. The witness node transmits
the vote directly to the base station. Forgery is not a problem in
this scheme. Moreover, fewer bits are necessary to represent the
vote, significantly reducing the power consumption. Performance
analysis and simulation results indicate that the proposed approach
can achieve a $40$ times better overhead than the witness-based
approach.
\end{abstract}


\newpage

\section{Introduction}
\label{sec:introduction} Wireless sensor networks (WSNs) comprise
many tiny, low-cost, battery-powered sensors in a small
area~\cite{akyildiz:wsn,culler:wsn,sohrabi:wsn_sa,alkaraki:wsn_routing,
sivrikaya:wsn_sync,niculescu:wsn_pos,niculescu:wsn_comm}. The
sensors detect environmental variations and then transmit the
detection results to other sensors or a base
station~\cite{aldosari:dd,chamberland:dd_power,tsitsiklis:dd,dan:dct,costa:dc}.
One or several sensors then collect the detection results from other
sensors. The collected data must be processed by the sensor to
reduce the transmission burden before they are transmitted to the
base station. This process is called {\it data fusion} and the
sensor performing data fusion is the {\it fusion node}. The fusion
data may be sent from the fusion node to the base station through
multiple hops~\cite{lin:multi-hops} or a direct
link~\cite{shen:sina}.

Although fusion significantly lowers the traffic between the fusion
node and the base station, the fusion node is more critical and
vulnerable to malicious attacks than non-fusion
sensors~\cite{sholander:assurance,perrig:wsn_security,olariu:ia}. If
a fusion node is compromised, then the base station cannot ensure
the correctness of the fusion data sent to it. This problem of
fusion data assurance arises because the detection results are not
sent directly to the base station, so the fusion result cannot
usually be verified.

This problem can be resolved in two ways: One is
hardware-based~\cite{anderson:tamper,olariu:virtual} and the other
is software-based~\cite{przydatek:aggregation,
deng:assurance,du:assurance}. Since the hardware-based approach
requires extra circuits to detect or frustrate the compromised node,
the cost and continual power consumption of sensors are increased
but still cannot guarantee protection against all attacks.
Conversely, the software methods generally require no or little
extra hardware for data assurance. However, as mentioned
in~\cite{przydatek:aggregation, deng:assurance}, several copies of
the fusion data must be sent to the base station, the power
consumption for the data transmission is very high.

The witness-based approach presented by Du {\it et
al.}~\cite{du:assurance} does not have this difficulty. Several
fusion nodes are used to fuse the collected data and have the
ability to communicate with the base station. Only one node is
chosen to transmit the fusion result to the base station. The other
fusion nodes, serving as witnesses, encrypt the fusion results to
message authentication codes (MACs). The MACs are then sent to the
base station through the chosen fusion node. Finally, the base
station utilizes the received MACs to verify the received fusion
data. The verification could be wrong since the chosen node could be
compromised and forge MACs. The correctness of the verification
depends not only on the number of malicious fusion nodes, but also
on the length of the MAC. A long MAC increases the reliability of
the verification. However, the transmission of the long MAC imposes
a high communication burden. If the received fusion result at the
base station cannot pass the verification, then a polling scheme is
started to determine whether any valid fusion result is available at
the other fusion nodes. In addition to the fusion result sent by the
malicious fusion node, several copies of the correct fusion result
may also have to be transmitted to the base station. The
transmission of the correct fusion result consumes the power of the
uncompromised fusion node.

This work develops a new data fusion assurance to improve the
witness-based method of Du {\it et al.}~\cite{du:assurance}. The
correctness of the verification in the proposed scheme depends only
on the number of compromised fusion nodes. As in the witness-based
approach, a fusion node is selected to transmit the fusion result,
while other fusion nodes serve as witnesses. Nevertheless, the base
station obtains votes contributing to the transmitted fusion result
directly from the witness nodes. No valid fusion data are available
if the transmitted fusion data are not approved by a pre-set number
of witness nodes. Based on this voting mechanism, two schemes are
addressed: one needs variant rounds of voting and the other needs
only one round of voting. In the variant-round scheme, only one copy
of the correct fusion data provided by one uncompromised fusion node
is transmitted to the base station. Analytical and simulation
results reveal that the proposed scheme is up to $40$ times better
on the overhead than that of Du {\it et al.}~\cite{du:assurance}. In
the one-round scheme, the base station polls each sensor once at
most. The maximum delay of the one-round scheme is much less than
the variant-round scheme.

The remainder of this work is organized as follows.
Section~\ref{sec:problem} briefly addresses the problem of data
fusion assurance in WSNs and previous works on the problem.
Section~\ref{sec:voting} describes the variant-round scheme and
analyzes its performance in terms of overhead and delay. The
description of the one-round scheme is also given.
Section~\ref{sec:performance} presents a performance evaluation of
the proposed approach. Concluding remarks and suggestions for future
work are presented in Section~\ref{sec:conclusions}.

\section{Data Fusion Assurance Problem and The Previous Work}
\begin{figure}
\centering
\includegraphics[width=4in]{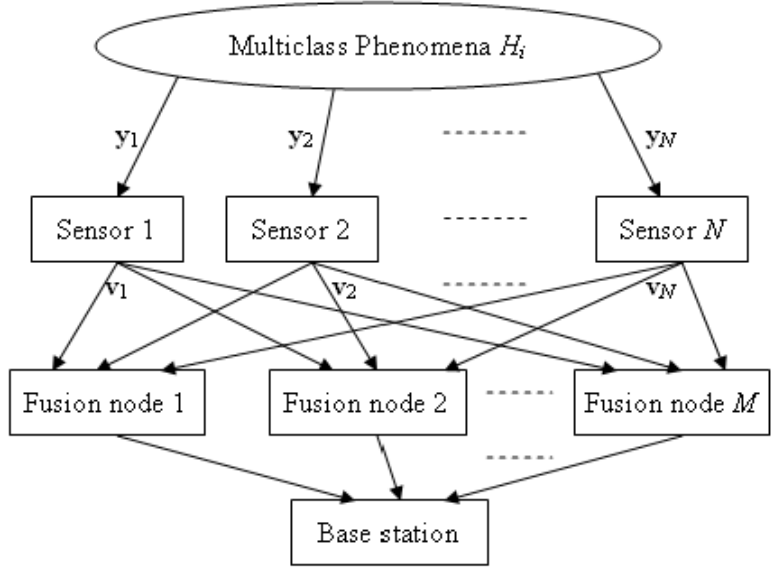}
\caption{Structure of a wireless sensor network for distributed
detection using $N$ sensors and $M$ fusion nodes}
\label{fig:structure}
\end{figure}
\label{sec:problem}Figure~\ref{fig:structure} depicts a wireless
sensor network for distributed detection with $N$ sensors for
collecting environment variation data, and a fusion center for
making a final decision of detections. This network architecture is
similar to the so-called SENsor with Mobile Access
(SENMA)~\cite{yang:misinformed,tong:senma}, Message
Ferry~\cite{zhao:ferrying}, and Data Mule~\cite{shah:mules}. At the
$j$th sensor, one observation $y_j$ is undertaken for one of
phenomena $H_i$, where $i=1,2,\ldots,L$. If the detection (raw) data
are transmitted to the fusion nodes without any processing, then the
transmission imposes a very high communication burden. Hence, each
sensor must make a local decision based on the raw data before
transmission. The decisions\footnote{These decisions could be
compressed data whose sizes depend on the applications of the
WSNs.}, $v_j$, $j=1,2,\ldots,N$, can be represented with fewer
symbols than the raw data. The sensor then transmits the local
decision to $M$ fusion nodes using broadcast. The fusion node can
combine all of the local decisions to yield a final result, and
directly communicate with the base station. Finally, one of the
fusion nodes is specified to send the final result to the base
station. Unless all of the fusion nodes or all of the sensors fail,
this detection and fusion scheme can guarantee that the base station
can obtain the detection result. However, the accuracy of the result
is not certain.

Two problems must be solved to ensure that the base station obtain
the correct result. First, every fusion node must correctly fuse all
of the local decisions, which also implies that all of the fusion
results must be the same. Several algorithms have been proposed to
deal with this
problem~\cite{varshney:dd,costa:dc,wang:dc_ft,wang:df_cc}. This work
assumes that this problem has been solved. The second problem
concerns assurance of the fusion result. The transmission between
the fusion node and the base station is assumed herein to be
error-free. Since some fusion nodes may be compromised, the fusion
node chosen by the base station to transmit the fusion result may be
one of the compromised nodes. Malicious data may be sent by the
compromised node, and the base station cannot discover the
compromised nodes from the normal fusion nodes since the data
detected by the sensor are not sent directly to the base station.
Consequently, the result obtained at the base station may be
incorrect.

Du {\it et al.}~\cite{du:assurance} presented a witness-based
approach to ensure the correctness of the fusion result. All fusion
nodes, other than the chosen node, acts as witnesses of the
transmitted fusion result. The witness nodes encrypt their own
fusion results to MACs with private keys shared with the base
station, and send the MACs, as "votes", to the chosen node. In the
$T+1$ out of $M$ voting scheme, the chosen node collects all MACs
from the witness nodes, and transmits them with its own fusion
result to the base station. The base station can determine the
received data whether the fusion result from the chosen node is
accurate. If the fusion result is supported by at least $T$ MACs,
then the base station accepts the fusion result. Normally,  $T >
\lfloor M/2 \rfloor$. However, although the number of compromised
nodes $C<T$, the accepted fusion result is not always correct. If
the chosen node is compromised, then it may forge the fusion result
and the MAC. Let the size of each MAC be $k_w$ (bits). Since the
number of the transmitted MACs is $M-1$, the number of the
transmitted bits, in addition to the fusion result, is $(M-1)k_w$.
The probability that the base station accepts the forged fusion
result is given by
\[
P_e = \sum_{i=T}^{M-1}\left(\begin{array}{c} M-1 \\ i
\end{array}\right)
\left(\frac{1}{2^{k_w}}\right)^i\left(1-\frac{1}{2^{k_w}}\right)^{M-i-1}.
\]
For instance, consider the majority voting rule in which $T+1 \ge
\lceil M/2 \rceil$. To ensure that $P_e \le 2^{-10}$, set
$k_w=\lceil 2\left(10/(M-1)+1\right)\rceil$. Additionally, although
only one copy of the fusion result is sent to the base station in
this witness-based approach, the witness nodes still requires
significant communication bandwidth because the MACs of the fusion
results are transmitted.

To compare the proposed scheme fairly with the witness-based
approach, the {\it overhead} is defined as the total number of bits,
except the bits for {\it one} copy of the correct fusion result,
transmitted to the base station by {\it uncompromised} nodes during
the data assurance process.\footnote{The method does not consider
the power consumption of all compromised nodes since they are not
useful to the WSN.} The overhead can therefore be regarded as the
{\it useful} power consumed for the data assurance. Moreover, the
{\it round delay} is defined as the number of rounds\footnote{Or the
number of fusion results sent to the base station.} required to
collect all MACs (votes) from the witness nodes and the {\it polling
delay} the number of votes (including all agree and disagree
voting).\footnote{The overall time delay is then can be derived by
these two delays.}

The overhead of the witness-based approach given
in~\cite{du:assurance} is then derived as follows. If the received
fusion result is not accepted, then the base station may start a
polling mechanism to seek the correct fusion result. The base
station randomly specifies another fusion node. The new chosen node
then sends its fusion result and all MACs from the witness nodes to
the base station.\footnote{All MACs must be sent to the base station
again to avoid the denial of service since the previously chosen
compromised fusion node might modify MACs it sent to the base
station. This action is not clearly presented
in~\cite{du:assurance}.} When the number of compromised fusion
nodes, $C$, is greater than $M-T-1$ but less than
$T+1$,\footnote{The $C$ compromised nodes are assumed to collude to
forge a wrong fusion result. Hence, if $C \ge T+1$, then they can
successfully forge a wrong fusion result, and the base station
accepts the forged result.} the fusion result is invalid and $M-T$
fusion nodes are chosen to transmit their fusion results to the base
station in the polling process. Since the overhead defined only
considers the power consumption of uncompromised nodes, the number
of uncompromised nodes among the $M-T$ fusion nodes must be
discovered. The probability that $i$ of the $M-T$ fusion nodes are
uncompromised is given by \setlength{\arraycolsep}{0.0em}
\begin{eqnarray*}
P_w(i)&{}={}&\frac{\left(\begin{array}{c} M-C \\
i
\end{array}\right) \left(\begin{array}{c} C \\M-T-i
\end{array}\right)} {\left(\begin{array}{c} M \\ M-T
\end{array}\right)} \\
&{}={}&\frac{\left(\begin{array}{c} M-T \\
i
\end{array}\right) \left(\begin{array}{c} T \\M-C-i
\end{array}\right)} {\left(\begin{array}{c} M \\ C
\end{array}\right)},
\end{eqnarray*}
\setlength{\arraycolsep}{5pt} where $0 \le i \le M-T$. Let $K$ be
the number of bits representing the fusion result. The average
overhead is thus
\begin{eqnarray}
\hspace{-0.2in} O_w(M,T,C) =  \sum_{i=1}^{M-T} P_w(i) \left[ (M-1)i
k_w + K(i-1) \right]  \mbox{ (bits)}, \label{eqn:witness_overhead}
\end{eqnarray}
where $i-1$ is used because the overhead is defined such that one
copy of correct fusion result does not count on. Equation
(\ref{eqn:witness_overhead}) indicates that the number of the
correct fusion results transmitted by the uncompromised fusion nodes
may be up to $M-T$. In other words, the power of the uncompromised
nodes is significantly wasted. Moreover, because each chosen node
has to collects all MACs from the witness nodes, the average round
delay, $R_w$, is $M-T$ and the average polling delay $D_w$ is
$(M-T)(M-1)$.

Conversely, when the number of uncompromised nodes is greater than
$T$, the base station can obtain the correct fusion result. If the
base station gets the correction result at round $i$, meaning that
the chosen fusion nodes from round $1$ to $i-1$ are compromised,
then the average round delay can be given by
\[
R_w = \sum_{i=1}^{C+1}i\frac{{\left(\begin{array}{c} M-i \\ C-i+1
\end{array}\right)}}{{\left(\begin{array}{c} M \\ C
\end{array}\right)}}
\]
and the average polling delay is $R_w(M-1)$. The average overhead is
given by $(M-1)k_w$. Notice that the maximum round delay is $C+1$.
When the fusion result is valid, $40$ and $60$ bits must be
transmitted to the base station when $M=11$ ($k_w = 4$, i.e., $P_e
\le 2^{-10}$) and $M=21$ ($k_w = 3$, i.e., $P_e \le 2^{-10}$),
respectively, setting $K=0$ (in practice, $K>0$). A large amount of
power must be consumed for this transmission, significantly reducing
the lifetime of the fusion node. The problem of power consumption is
even worse when the fusion result is invalid. For example, the
maximum average overhead is about $109$ and $314$ bits for $M=11$
($C=5$ and $T=6$) and $M=21$ ($C=10$ and $T=11$), respectively.
Therefore, the witness-based approach must be enhanced.

\section{Improved Voting Mechanism}
\label{sec:voting} The voting mechanism in the witness-based
approach is designed according to the MAC of the fusion result at
each witness node. This design is reasonable when the witness node
does not know about the fusion result at the chosen node. However,
in practice, the base station can transmit the fusion result of the
chosen node to the witness or the witness node is in the
communication range of the chosen node and the base station.
Therefore, the witness node can obtain the transmitted fusion result
from the chosen node through the base station or overhearing. The
witness node then can compare the transmitted fusion result with its
own fusion result. Finally, the witness node can send its vote
(agreement or disagreement) on the transmitted result directly to
the base station, rather than through the chosen node.

The base station has to set up a group key for all fusion nodes to
ensure that the direct voting mechanism works.\footnote{We assume
that all witnesses can overhear the fusion result sent by the chosen
node. If it is not the case, the direct voting mechanism needs to be
slightly modified.}  When a fusion node wishes to send its fusion
result to the base station, it adopts the group key to encrypt the
result, and other fusion nodes serving as witness nodes can decode
the encrypted result. The witness node then starts to vote on the
transmitted result. Two data fusion assurance schemes are proposed
based on the voting mechanism using a group key.

\subsection{Variant-round Scheme}
\label{subsec:variant} In this scheme, the base station needs to ask
the witness node whether it agrees or disagrees with the transmitted
fusion result. The witness node then sends its vote to the base
station. If the transmitted fusion result is not supported by at
least $T$ witness nodes, then the base station might have to select
a witness node that does not agree with the transmitted result as
the next chosen node. The detail steps of the scheme are given as
follows:
\begin{description}
\item[Step 1: ]\ The base station chooses a fusion node.
Other fusion nodes serve as witness nodes. Define a set of witness
nodes that includes all witness nodes and let the nodes in the set
be randomly ordered. Denote $M^{\prime} = M-1$ as the size of the
witness set in the current round.

\item[Step 2: ]\ The chosen node transmits its fusion result to
the base station.

\item[Step 3: ]\ The base station polls the node in the witness set by
following the order of the witness nodes. The
polling-for-vote\footnote{The rest of this work utilizes "voting",
"polling" and "polling-for-vote" interchangeably.} process does not
stop until
\begin{itemize}
 \item $T$ witness nodes {\it agree} with the transmitted fusion result (agreeing nodes), where $1 \le T
       \le M-1$,
 \item $M^{\prime}-T+1$ witness nodes {\it disagree} with the transmitted fusion
 result (disagreeing nodes), or
 \item all witness nodes have been polled.
\end{itemize}
\item[Step 4: ]\ Represent $A$ as the number of witness nodes that agree
with the transmitted fusion result.
\begin{itemize}
 \item If $A = T$, then the transmitted fusion result passes the
 verification of the fusion result. Stop the polling.
 \item If $M^{\prime}-T-1<A<T$, then no reliable fusion result is valid.
 Stop the polling.
 \item If $A \le M^{\prime}-T-1$, then exclude the $A$ agreeing witness nodes
from the witness set. Let the first node that disagrees with the
transmitted fusion result be the chosen node to transmit its fusion
result. Thus, the updated size of the witness set, $M^{\prime}$,
becomes to $M^{\prime}-A-1$\footnote{The number of nodes performing
the polling-for-vote at next round becomes to $M^{\prime}-A$.}. Go
to Step~2 for the next round of the polling.
\end{itemize}
\end{description}

\subsection{Analysis of the Variant-round Scheme}
\label{sec:analysis_VR}

This analysis assumes that the compromised node always transmits the
forged fusion result while the compromised node is chosen to send
its fusion result. When a compromised node serves as a witness node,
it always disagrees with the correct fusion result, and agrees with
the forged fusion result with a probability $P_f$. If the
compromised node attempts to make the base station accept the forged
fusion result, then it always agrees with the fusion result
transmitted by other compromised nodes, i.e., $P_f=1$, and at most
two rounds of voting have to be run. Conversely, if the compromised
node wants to make the polling-for-vote process run as long as
possible, then it always disagrees with the transmitted fusion
result, i.e., $P_f=0$. The performance of the variant-round scheme
when $P_f=0$ is analyzed next. Appendix describes the performance
analysis of the variant-round scheme when $P_f=1$. When $P_f \ne
0,1$, the overheads are presented by computer simulations in next
section. Since all compromised nodes (uncompromised nodes) have the
same behavior in the following analysis, the analysis can be treated
as the problem of counting for $C$ black balls (compromised nodes)
and $M-C$ white balls (uncompromised nodes) together.

If the compromised node always disagrees with the transmitted fusion
result, then no forged fusion result is accepted. Two cases must be
addressed:
\begin{description}
\item[Case 1]\ $C \ge M-T$,
\item[Case 2]\ $C<M-T$.
\end{description}
Note that the valid fusion result is not available in Case $1$.

Assume that the chosen node at the first round {\bf is} compromised.
The probability that the chosen node is compromised at the first
round is given by $C/M$. The first-round voting finishes when $M-T$
witness nodes do not agree with the transmitted fusion result, as
described in Step~3. Thus, the polling order (i.e., the order of
witness nodes as described in Step~1) determines the number of
witness nodes that the base station has to poll at this round of
voting. The number of possible polling orders in the sense of the
black-white-ball model is given by
\[
\Pi_{v1}^{c1} = \frac{(M-1)!}{(C-1)!(M-C)!} = \left(\begin{array}{c} M-1 \\
C-1
\end{array}\right),
\]
where the subscript, $v1$, denotes the first case of the
variant-round scheme, and the superscript, $c1$, represents the
first round of voting when the chosen node is compromised. Since the
chosen node at the first-round voting is compromised, the polling
stops after $M-T$ witness nodes have been polled. Moreover, the
number of unpolled nodes is $M-1-(M-T)=T-1$, and the number of
uncompromised nodes among the unpolled nodes is $M-C-i$ if there are
$i$ uncompromised nodes polled. Thus, the probability that $i$ of
the $M-T$ polled nodes are uncompromised, where $0\le i \le
M-T$,\footnote{Actually, $\max \{M-T-C+1,0\}\le i$. Since ${a
\choose b}$ is $0$ when $b>a$, $0$ is adopted as the lower bound of
$i$.} is then written by
\[
\frac{1}{\Pi_{v1}^{c1}}{\left(\begin{array}{c} M-T \\ i
\end{array}\right) \left(\begin{array}{c} T-1 \\M-C-i
\end{array}\right)}.
\]
No node is excluded from the witness set. The number of compromised
nodes and the size of the witness set at the second round become
$C-1$ and $M-2$, respectively. The average overhead, when the chosen
node at the first round is compromised, is expressed recursively by
\begin{equation}
O_{v1}^{c}(M,T,C,0) = \sum_{i=0}^{M-T}\frac{1}{\Pi_{v1}^{c1}}
{\left(\begin{array}{c} M-T
\\ i \end{array}\right) \left(\begin{array}{c} T-1 \\M-C-i
\end{array}\right)} i
k^{\prime} + O_{v1}(M-1,T,C-1,0),
\label{eqn:polling_invalid_compromised_overhead0}
\end{equation}
where $O_{v1}(M-1,T,C-1,0)$ represents the average overhead when the
number of fusion nodes is $M-1$ and the number of compromised nodes
is $C-1$, and $k$ ($k^{\prime}$) is the number of bits that a
witness must send to the base station while it agrees (disagrees)
with the transmitted fusion result.\footnote{The bits sent to the
base station when a node agrees with the fusion result are separated
from those sent when the node disagrees with the result, since the
node can be silenced when it agrees with the result, while only a
few bits are sent when it disagrees.} Moreover, the average round
delay and the average polling delay under the same condition are
represented by
\begin{equation}
R_{v1}^c(M,T,C,0) = 1 + R_{v1}(M-1,T,C-1,0),
\label{eqn:polling_invalid_compromised_round_delay0}
\end{equation}
and
\begin{equation}
D_{v1}^c(M,T,C,0) = M-T + D_{v1}(M-1,T,C-1,0),
\label{eqn:polling_invalid_compromised_polling_delay0}
\end{equation}
where $R_{v1}(M-1,T,C-1,0)$ and $D_{v1}(M-1,T,C-1,0)$ are the
average round delay and polling delay, respectively, when the number
of fusion nodes performing the polling-for-vote is $M-1$ and the
number of compromised nodes among them is $C-1$.

Next, suppose that the chosen node at the first round {\bf is  not}
compromised, which has a probability of $(M-C)/M$. The number of the
possible polling orders is given by
\[
\Pi_{v1}^{u1} = \left(\begin{array}{c} M-1 \\ C \end{array}\right),
\]
where the superscript, $u1$, denotes the first round of polling
while the chosen node is uncompromised. When the polling stops at
witness node $j$, the node does not agree with the transmitted
result and the base station has polled $M-T$ disagreeing nodes
(including witness node $j$). Moreover, $M-j-1$ nodes are unpolled,
and $C-(M-T)=T+C-M$ of these are compromised. Since the witness set
has $M-C-1$ uncompromised nodes, the maximum number of polled
witness nodes is $(M-C-1)+(M-T)$. Thus, the probability that the
polling stops at the $j$th witness node, where $M-T \le j \le
(M-C-1)+(M-T)=2M-T-C-1$, is given by
\begin{equation*}
P_{v1}^{u1}(j)=\frac{1}{\Pi_{v1}^{u1}}\left(\begin{array}{c} j-1 \\
M-T-1
\end{array}\right) \left(\begin{array}{c} M-j-1 \\ T+C-M
\end{array}\right).
\end{equation*}

Since the number of unpolled nodes $M-j-1$ is less than $T$, these
nodes will never be polled in the following runs before the voting
mechanism stops. Hence, only compromised nodes are polled after the
first round  and then no uncompromised nodes are further polled.
Note that the number of uncompromised nodes among the polled nodes
is $j-(M-T)=j-M+T$. Therefore, the average overhead, when the chosen
node at the first round {\bf is not} compromised, is then given by
\begin{equation}
O_{v1}^u(M,T,C,0) = \sum_{j=M-T}^{2M-T-C-1}P_{v1}^{u1}(j)(j-M+T)k.
\label{eqn:polling_invalid_uncompromised_overhead0}
\end{equation}
Moreover, the size of the witness set after the first round becomes
$M^{\prime}=(M-j-1)+(M-T)-1=2M-T-j-2$ (the total number of unpolled
and disagreeing nodes minus 1). The polling process stops if
\begin{eqnarray*}
&&M^{\prime} \le T-1 \\
& \Leftrightarrow & 2M-T-j-2 \le T-1 \\
& \Leftrightarrow & j \ge 2M-2T-1.
\end{eqnarray*}
Otherwise, the size is decreased by $1$ in each following round
until it becomes $T$. Consequently, the total number of the
following rounds is $2M-T-j-2-T+1=2M-2T-j-1$ and then the number of
the total rounds is $2M-2T-j-1+1=2M-2T-j$. The average round delay
is then represented by
\begin{equation}
R_{v1}^u(M,T,C,0) =
\sum_{j=M-T}^{2M-2T-2}P_{v1}^{u1}(j)(2M-2T-j)+\sum_{j=2M-2T-1}^{2M-T-C-1}P_{v1}^{u1}(j).
\label{eqn:polling_invalid_uncompromised_round_delay0}
\end{equation}
Since the voting process will stop when $M^{\prime}-T+1$ nodes are
polled at each round, the polling delay is
\begin{eqnarray}
D_{v1}^u(M,T,C,0) &=&
\sum_{j=M-T}^{2M-T-C-1}P_{v1}^{u1}(j)j\nonumber\\
&+&\sum^{2M-2T-2}_{j=M-T}P_{v1}^{u1}(j)\frac{(2M-2T-j)(2M-2T-j-1)}{2}.
\label{eqn:polling_invalid_uncompromised_polling_delay0}
\end{eqnarray}

 Equations
(\ref{eqn:polling_invalid_compromised_overhead0}) to
(\ref{eqn:polling_invalid_uncompromised_polling_delay0}), and the
initial conditions then give the average overhead and the average
delays of Case 1 as
\begin{eqnarray*}
&&O_{v1}(M,T,C,0) = \left\{\begin{array}{ll} 0 & M \le T \\
\frac{C}{M}O_{v1}^c(M,T,C,0)+\frac{M-C}{M}O_{v1}^u(M,T,C,0) & \mbox{
else}
\end{array}, \right. \\
&&R_{v1}(M,T,C,0) = \left\{\begin{array}{ll} 0 & M \le T \\
\frac{C}{M}R_{v1}^c(M,T,C,0)+\frac{(M-C)}{M}R_{v1}^u(M,T,C,0) &
\mbox{ else}
\end{array}, \right.
\\
&&D_{v1}(M,T,C,0) = \left\{\begin{array}{ll} 0 & M \le T \\
\frac{C}{M}D_{v1}^c(M,T,C,0)+\frac{(M-C)}{M}D_{v1}^u(M,T,C,0) &
\mbox{ else}
\end{array}. \right.
\end{eqnarray*}

The second case, i.e., $C<M-T$, produces a valid fusion result.
Similarly, if the chosen node at the first-round polling is
compromised, then the polling stops after $M-T$ witness nodes are
polled. The average overhead and the average delays, when the chosen
node at the first round is compromised, are expressed respectively
as
\begin{eqnarray*}
&&O_{v2}^c(M,T,C,0)=\sum_{i=M-T-C+1}^{M-T}\frac{1}{\Pi_{v1}^{c1}}{\left(\begin{array}{c}
M-T
\\ i
\end{array}\right) \left(\begin{array}{c} T-1 \\M-C-i
\end{array}\right)}i k^{\prime}+ O_{v2}(M-1,T,C-1,0),\\
&&R_{v2}^c(M,T,C,0)=1+ R_{v2}(M-1,T,C-1,0),\\
&&D_{v2}^c(M,T,C,0)=M-T+ D_{v2}(M-1,T,C-1,0).
\end{eqnarray*}
Only one round of polling is needed when the chosen node is
uncompromised at the first round. When the polling stops at witness
node $j$, the node agrees with the transmitted result and the base
station has polled $T$ agreeing nodes (including witness node $j$).
Moreover, $M-j-1$ nodes are unpolled, and $M-1-C-T$ of these are
uncompromised. The probability that the polling process ends at the
$j$th witness node, where $T \le j \le T+C$, is given by
\begin{equation*}
P_{v2}^{u1}(j) = \frac{1}{\Pi_{v1}^{u1}}\left(\begin{array}{c} j-1 \\
T-1
\end{array}\right) \left(\begin{array}{c} M-j-1 \\ M-C-T-1
\end{array}\right).
\end{equation*}
The number of polled uncompromised nodes is $T$. The average
overhead when the chosen node is uncompromised at the first round is
represented as
\begin{equation*}
O_{v2}^u(M,T,C,0)=\sum_{j=T}^{T+C} P_{v2}^{u1}(j) Tk,
\end{equation*}
and the average polling delay is
\begin{equation*}
D_{v2}^u(M,T,C,0)=\sum_{j=T}^{T+C} P_{v2}^{u1}(j) j.
\end{equation*}
 Consequently, the average
overhead $O_{v2}(M,T,C,0)$, the average round delay
$R_{v2}(M,T,C,0)$, and the average polling delay $D_{v2}(M,T,C,0)$
can be represented as
\begin{eqnarray*}
&&O_{v2}(M,T,C,0) = \left\{\begin{array}{ll} 0 & M \le T \\
Tk & M>T
\mbox{ and } C = 0 \\
\frac{C}{M}O_{v2}^c(M,T,C,0)+\frac{M-C}{M}O_{v2}^u(M,T,C,0) & \mbox{
else}
\end{array} \right. , \\
&&R_{v2}(M,T,C,0) = \left\{\begin{array}{ll} 0 & M \le T \\
1 & M>T
\mbox{ and } C = 0 \\
\frac{C}{M}R_{v2}^c(M,T,C,0)+\frac{M-C}{M} & \mbox{ else}
\end{array} \right. ,\\
&&D_{v2}(M,T,C,0) = \left\{\begin{array}{ll} 0 & M \le T \\
T & M>T
\mbox{ and } C = 0 \\
\frac{C}{M}D_{v2}^c(M,T,C,0)+\frac{M-C}{M}D_{v2}^u(M,T,C,0) & \mbox{
else}
\end{array} \right. .
\end{eqnarray*}

An interesting property of the variant-round scheme is that
throughout the polling process, at most one fusion result is
transmitted from all of the uncompromised nodes to the base station.
Hence, the overhead of the scheme is independent of the size of the
fusion result.\footnote{Note that the overhead is defined as the
total number of bits transmitted minus the bits of the correct
fusion result.} This claim can be proven by the following argument.
In the case of a valid fusion result, when the fusion result from an
uncompromised node is sent to the base station at a round of
polling, the polling process stops at this round and the valid
fusion result is obtained by the base station. Accordingly, only one
valid fusion result is sent to the base station by all uncompromised
nodes. In the case of no valid fusion result, the number of
uncompromised nodes as witnesses is less than $T$. If a round of the
polling process is the first time that the chosen node is
uncompromised, this round terminates when either all witness nodes
have been polled or $M'-T+1$ witness nodes disagree with the
transmitted fusion result. In the former case, the polling process
stops. In the latter case, another polling round is required.
Importantly, in the next round all uncompromised nodes will be the
last $T-1$ nodes in the witness set and will not then be chosen to
send any fusion result before the polling process is
completed.\footnote{In the next round, all uncompromised polled
nodes are deleted from the witness set according to the scheme.}
Therefore, only one fusion result is sent by all uncompromised nodes
when no valid result can be obtained by the base station.

\subsection{One-round Scheme} The number of rounds in the above scheme is not
fixed. Hence, the delay varies. Variant delay is not desired in some
applications such as real-time systems. This work proposes another
scheme that is based on the improved voting mechanism. In this
scheme the base station may receive different fusion results from
the witness nodes. It requires that all received fusion results be
stored. This scheme has a fixed delay and is summarized  as follows:
\begin{description}
\item[Step 1: ]\ The base station randomly chooses a fusion node.
Other fusion nodes serve as witness nodes. Define a set of witness
nodes that includes all of the witness nodes and let the nodes in
the set be randomly ordered.

\item[Step 2: ]\ The chosen node transmits its fusion result to
the base station. The base station sets the fusion result as a
potential {\bf voting} result and the number of agreeing votes for
the fusion result is set to be zero.

\item[Step 3: ]\ The base station polls the nodes, with the voting
result, one by one in the witness set, following the order of the
witness nodes. The witness node compares its fusion result with the
voting result.
\begin{itemize}
 \item If the witness node agrees with the voting result, it sends
an agreeing vote to the base station. The base station increases the
number of agreeing votes for the voting result by one.
 \item If the witness node does not agree with the voting result,
it transmits its fusion result to the base station.
  \begin{itemize}
   \item If the fusion result has been stored in the base station, then the
    base station increases the number of agreeing votes for the fusion result
    by one.
   \item If the fusion result has not been stored in the base station,
    then the base station stores the fusion result and the number of agreeing
    votes for the fusion result is set to be zero.
  \end{itemize}
\end{itemize}
The base station sets the voting result to the received fusion
result with the maximum number of agreeing votes to poll the next
witness node. If two or more fusion results received the maximum
umber of votes, then the voting result is set to the result with the
most recent vote. The polling stops when any received fusion result
receives $T$ votes or when the number of unpolled nodes plus the
maximum number of votes for the results recorded at the base station
is less than $T$.
\end{description}

\section{Performance Evaluation}
\label{sec:performance} In this section numerical and computer
simulations are performed to evaluate the performance of the
proposed schemes. The performances of the proposed variant-round
scheme are numerically calculated by the results given in
Section~\ref{sec:voting} when $P_f=0$ and $1$. The performances of
the variant-round scheme when $P_f\neq 0$ or $1$ and that of the
one-round scheme are evaluated using Monte Carlo computer
simulations. The proposed schemes are compared using the
witness-based approach in terms of overhead, average round delay,
and average polling delay performances. For the witness-based
approach, the size of each MAC, $k_w$, is assumed to be four bits.
In the evaluation of the overhead of the variant-round scheme, the
size of fusion result is zero (i.e., $K=0$), such that the
witness-based approach has the best performance. As stated at the
end of Subsection~\ref{sec:analysis_VR}, the overhead of the
variant-round scheme is independent of the size of the fusion
result. Under this assumption, the variant-round scheme is inferred
to have better overhead performance than the witness-based approach
for all sizes of the fusion result whenever it outperforms with a
zero-sized the fusion result. All results are presented for number
of nodes $M=11$.\footnote{Similar results can be obtained for
$M=21$, but omitted because of the page limit.}

\begin{figure}
\centering
\includegraphics[width=4in]{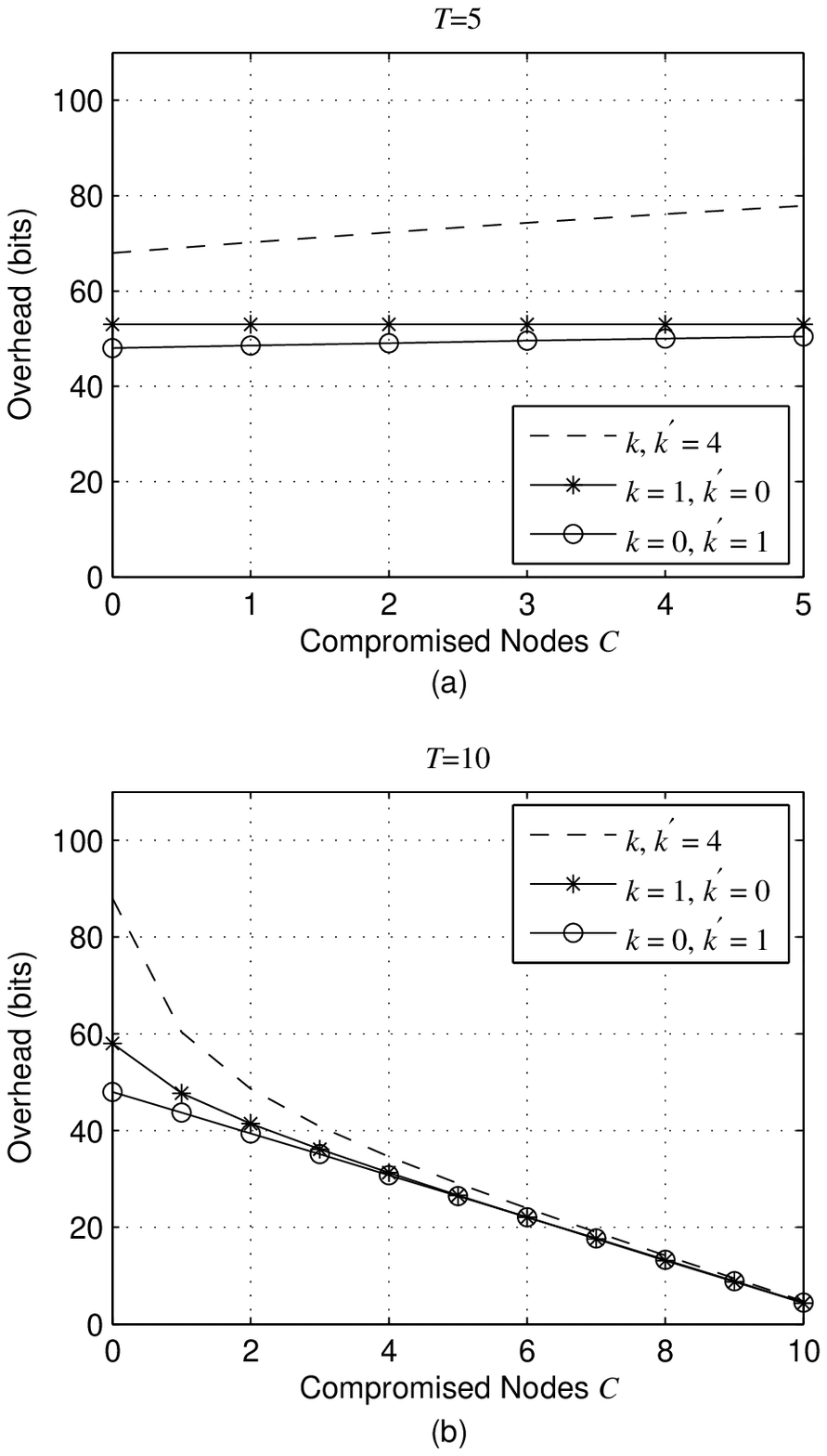}
\caption{The overheads of the variant-round scheme, for $M=11$ and
$P_f=0$, (a) when $T=5$ (valid fusion result) and (b) when $T=10$
(invalid fusion result).} \label{fig:v_analysis11_0}
\end{figure}

\begin{figure}
\centering
\includegraphics[width=4in]{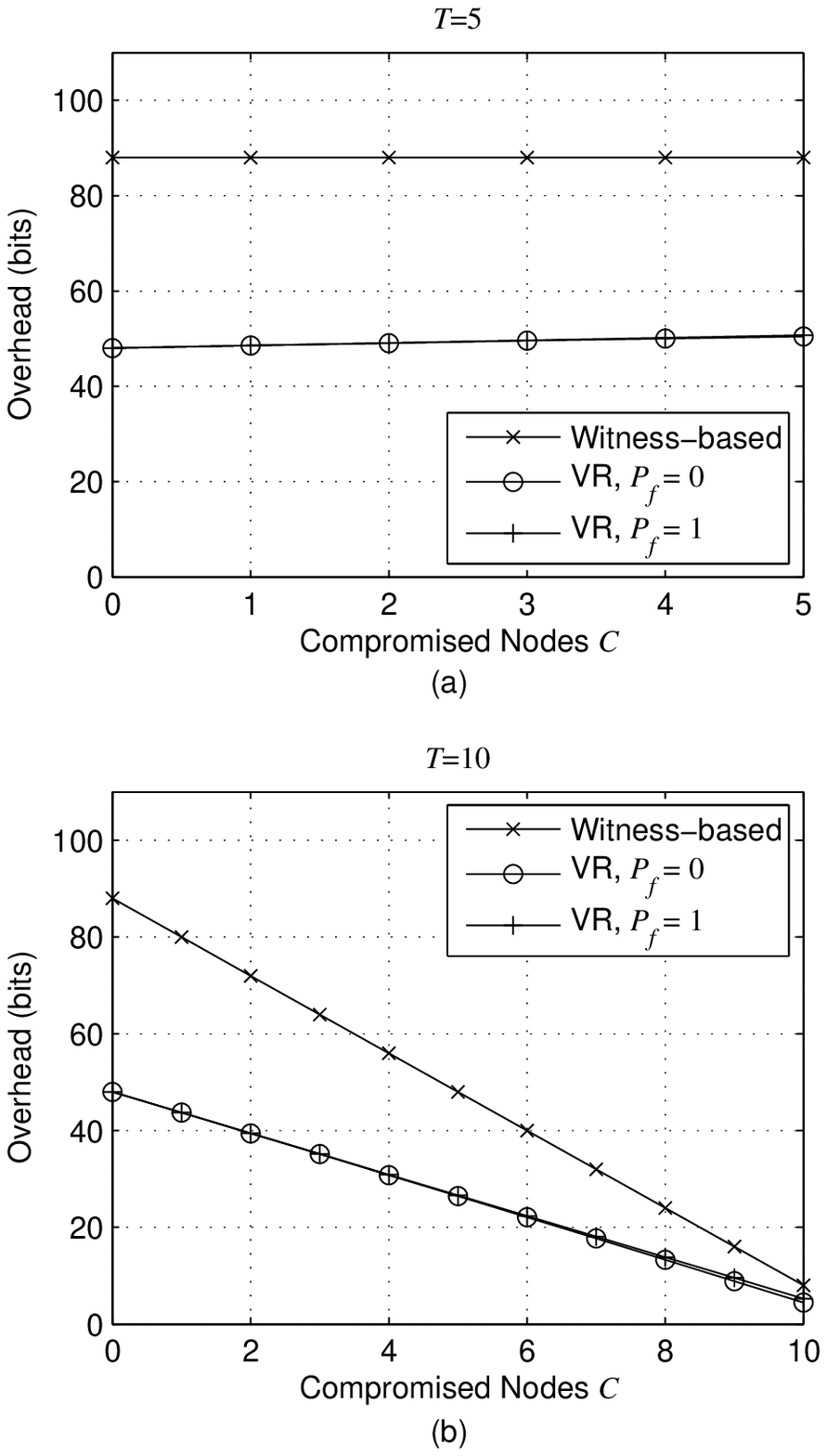}
\caption{Overhead comparison between the variant-round (VR) scheme
and the witness-based approach given in~\cite{du:assurance}, for
$M=11$ and $P_f=1,0$, (a) when $T=5$ (valid fusion result) and (b)
when $T=10$ (invalid fusion result).} \label{fig:analysis_comp11}
\end{figure}

\begin{figure}
\centering
\includegraphics[width=4in]{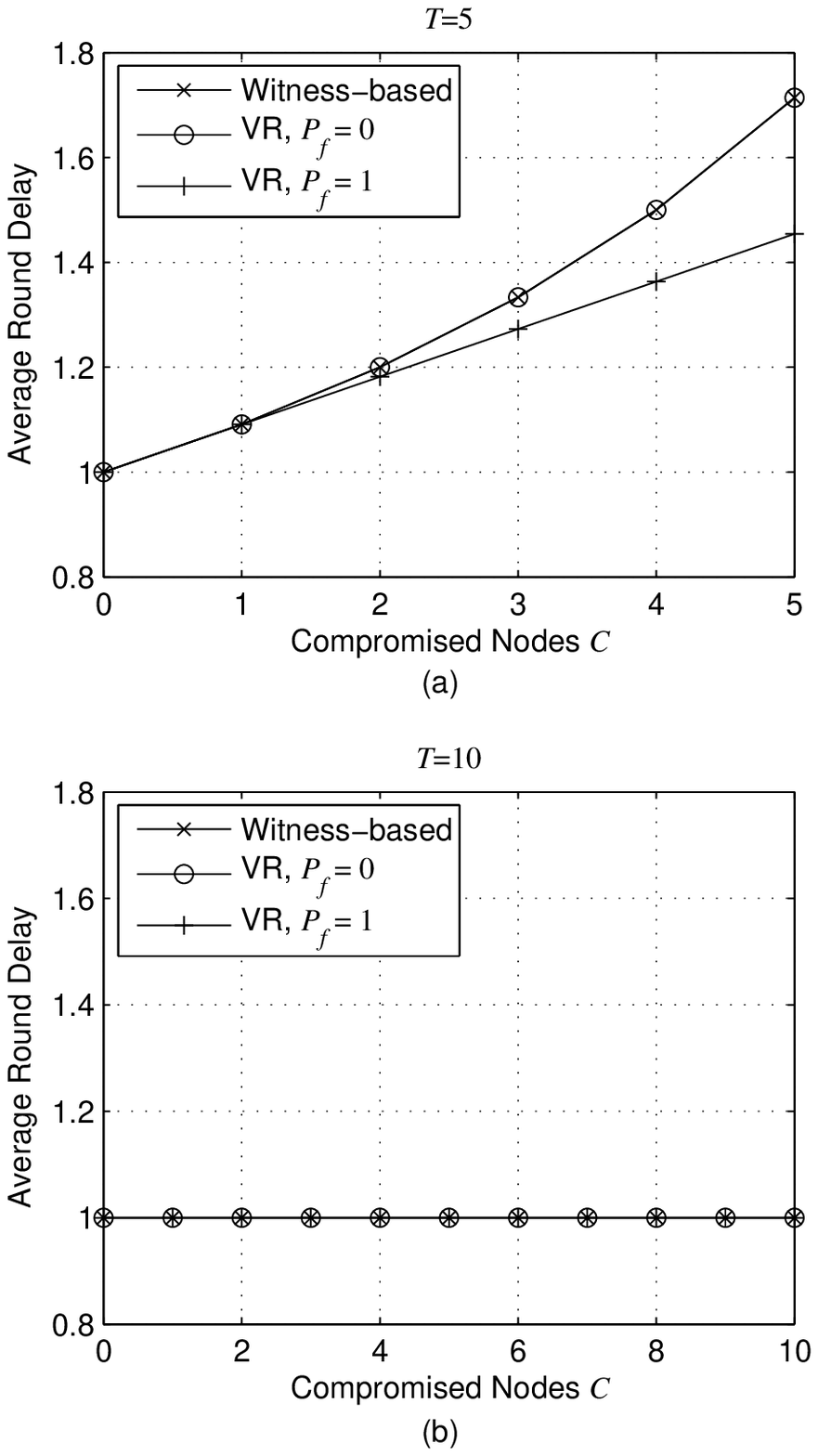}
\caption{Average round delay comparison between the variant-round
(VR) scheme and the witness-based approach given
in~\cite{du:assurance}, for $M=11$ and $P_f=1,0$, (a) when $T=5$
(valid fusion result) and (b) when $T=10$ (invalid fusion result).}
\label{fig:round_analysis_comp11}
\end{figure}

\begin{figure}
\centering
\includegraphics[width=4in]{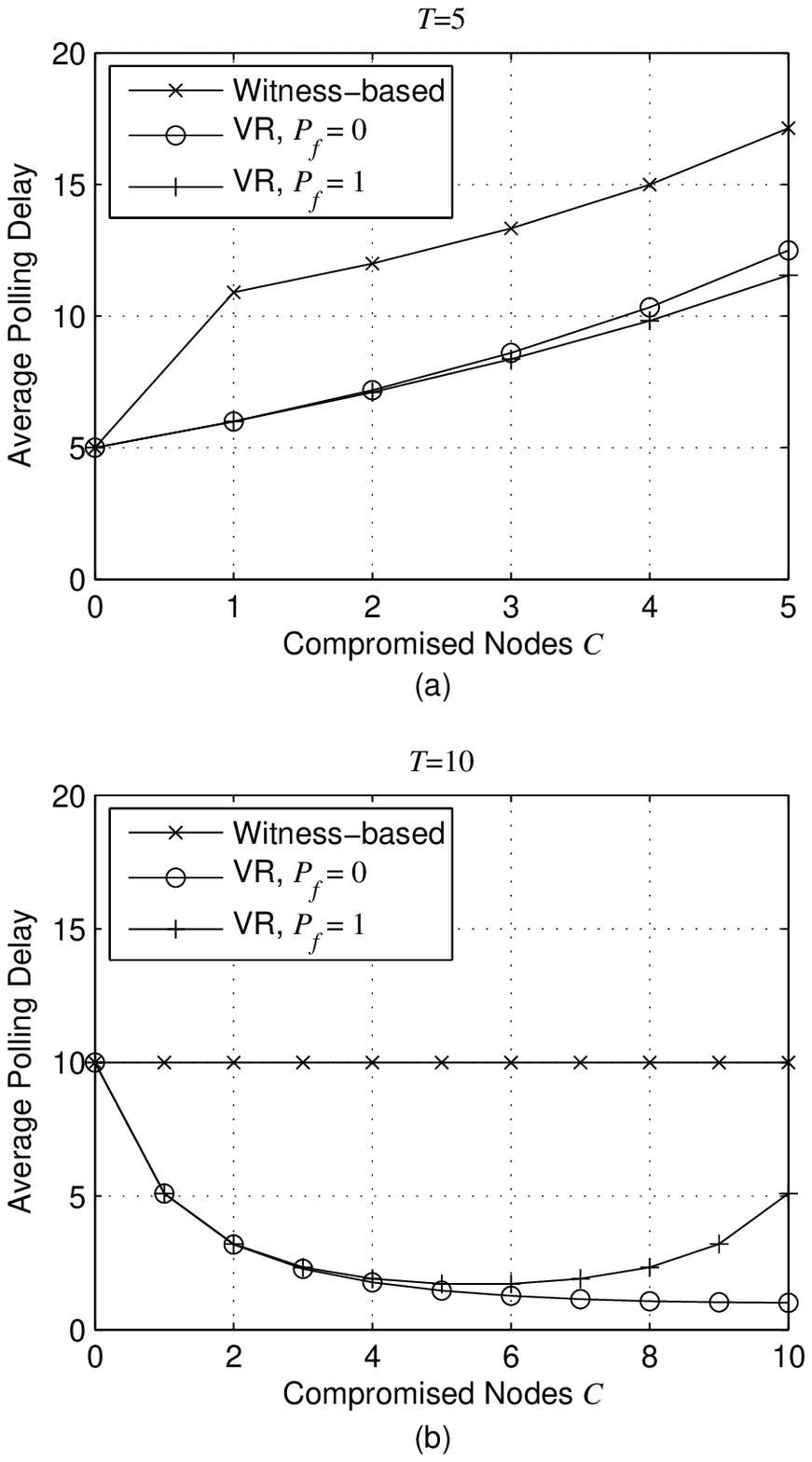}
\caption{Average polling delay comparison between the variant-round
(VR) scheme and the witness-based approach given
in~\cite{du:assurance}, for $M=11$ and $P_f=1,0$, (a) when $T=5$
(valid fusion result) and (b) when $T=10$ (invalid fusion result).}
\label{fig:delay_analysis_comp11}
\end{figure}

Figure~\ref{fig:v_analysis11_0} shows the overheads of the
variant-round scheme  when $P_f=0$. As in the example in
Section~\ref{sec:problem}, $k,k^{\prime}=4$ is set first and then
set $k=1,k^{\prime}=0$ and $k=0,k^{\prime}=1$ are considered. This
figure indicates that setting $k=1$ and $k^{\prime}=0$ significantly
reduces the overhead of the proposed scheme, independently of the
fusion result at the base station. Setting $k=0$ and $k^{\prime}=1$
further reduces the overhead of the proposed scheme. Next, the
variant-round (VR) scheme is compared with the witness-based
approach. In the variant-round scheme, $k=0$ and $k^{\prime}=1$ are
set. Figure~\ref{fig:analysis_comp11} compares overheads. The
variant-round scheme significantly outperforms the witness-based
approach regardless of the fusion result at the base station. For
example, according to Fig.~\ref{fig:analysis_comp11}, when $T=5$ and
$M=11$ the variant-round scheme is almost $40$ times better than the
witness-based approach given in~\cite{du:assurance} in terms of
overhead performance for $C=1$ and $2.$
Figure~\ref{fig:round_analysis_comp11} compares the average round
delays of the proposed variant-round scheme and the witness-based
approach. This figure demonstrates that the average round delays of
the proposed scheme are smaller than those of the witness-based
approach when the base station can obtain the valid fusion results;
however, they perform equally when the base station obtains invalid
fusion results. Figure~\ref{fig:delay_analysis_comp11} compares the
average polling delays of the proposed variant-round scheme and the
witness-based approach. In Fig.~\ref{fig:delay_analysis_comp11}, the
proposed scheme has much smaller average polling delays than the
witness-based approach - unlike average round delay performance. For
example, when $T=10$ and $M=11$, the proposed variant-round scheme
is almost five times better than the witness-based approach in terms
of average polling delay performance for $C=4$ and $5.$ Accordingly,
the proposed scheme outperforms the witness-based approach given
in~\cite{du:assurance} in terms of overhead and delay.


\begin{figure}
\centering
\includegraphics[width=4in]{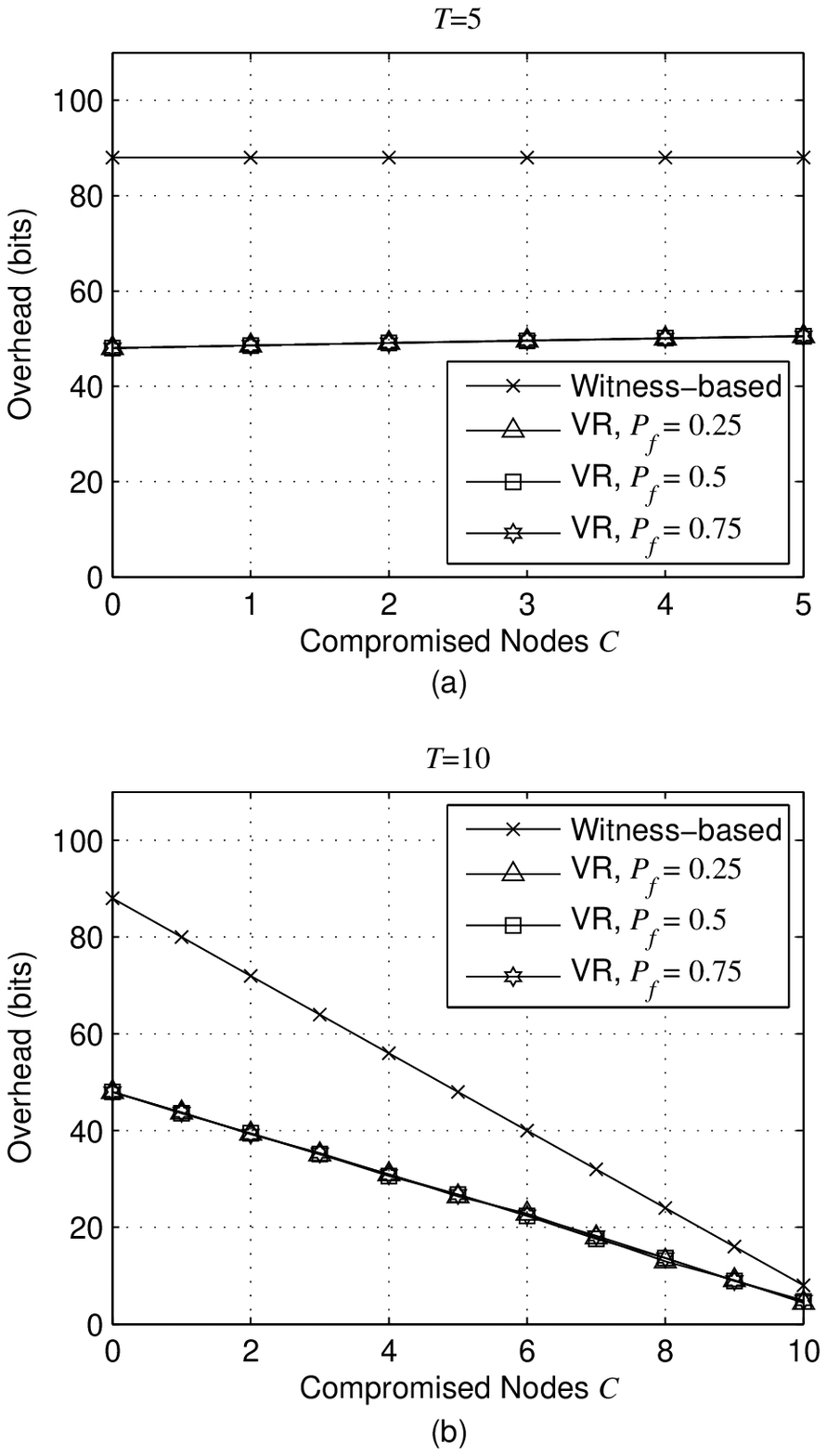}
\caption{Overhead comparison between the variant-round (VR) scheme
and the witness-based approach given in~\cite{du:assurance}, for
$M=11$ and $P_f=0.25,0.5,0.75$, (a) when $T=5$ (valid fusion result)
and (b) when $T=10$ (invalid fusion result).}
\label{fig:numerical_comp11}
\end{figure}

The following computer simulations evaluate the variant-round scheme
when $P_f = 0.25, 0.5,$ and $0.75$ by performing $10000$ Monte Carlo
tests for each simulation. In the first set of simulations, the
witness-based approach given in~\cite{du:assurance} is compared with
the proposed variant-round scheme when $P_f = 0.25, 0.5,$ and
$0.75$. Figure~\ref{fig:numerical_comp11} presents the results for
$M=11$, $k=0$ and $k^{\prime}=1$. The variant-round scheme
outperforms the witness-based approach in every $P_f$ simulated. For
example, in Fig.~\ref{fig:numerical_comp11}, when $T=5$ the proposed
variant-round scheme is almost $40$ times better than the
witness-based approach given in~\cite{du:assurance} in terms of
overhead performance for $C=1,2$ and $P_f=0.25, 0.5, 0.75$. Finally,
Table~\ref{tbl:max_overhead_vr} summarizes the maximum average
overhead of the variant-round scheme and the witness-based approach
for $M=11$ and $12$.

\begin{table}
\centering
\renewcommand{\arraystretch}{1.3}
\caption{Maximum average overheads (bits) for variant-round (VR)
scheme and witness-based approach given
in~{\protect\cite{du:assurance}}} \centering
\begin{tabular}{|c||c|c|c|c|c|c|}
\hline
$$ & Witness-based & VR, $P_f=0$ & VR, $P_f=0.25$ & VR, $P_f=0.5$ & VR, $P_f=0.75$ & VR, $P_f=1$ \\
\hline
$M=11$ & 109 & 2.5 & 2.5 & 2.4 & 2.7 & 3.2\\
\hline
$M=21$ & 314 & 4.9 & 4.7 & 4.7 & 4.9 & 6.2 \\
\hline
\end{tabular} \label{tbl:max_overhead_vr}
\end{table}

\begin{figure}
\centering
\includegraphics[width=4in]{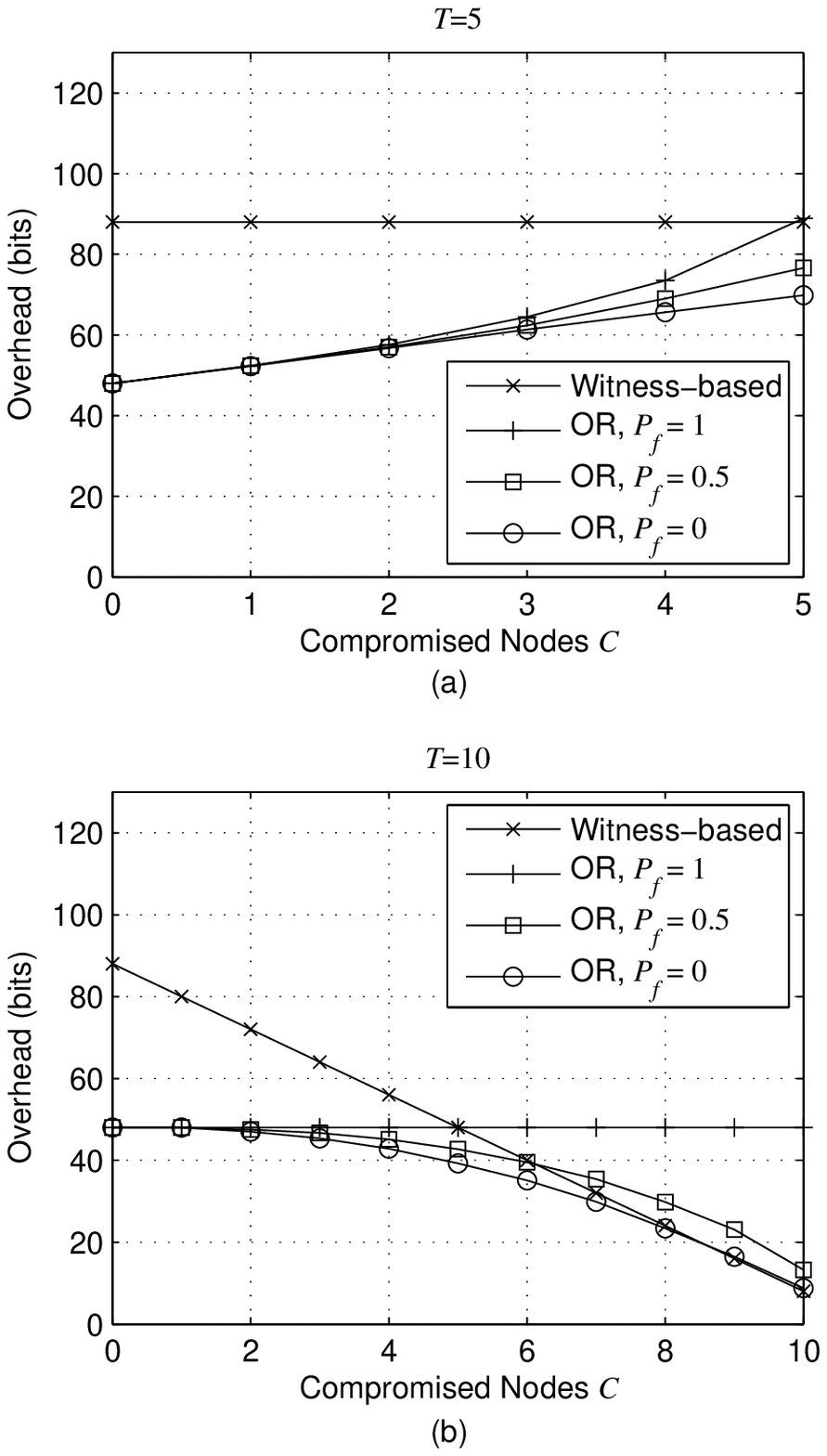}
\caption{Overhead comparison between the one-round (OR) scheme and
the witness-based approach given in~\cite{du:assurance}, for $M=11$
and $P_f=0,0.5,1$, (a) when $T=5$ (valid fusion result) and (b) when
$T=10$ (invalid fusion result).} \label{fig:oneround11}
\end{figure}

In the second set of simulations, the average numbers of bits sent
by uncompromised nodes in the one-round scheme is evaluated. When
the compromised node does not agree with the voting result, the
fusion result transmitted by the compromised node is different from
other fusion results and the size of the fusion result is $K=48$.
Fig.~\ref{fig:oneround11} shows the results for $M=11$ when $P_f=0,
0.5$, and $1$. In Fig.~\ref{fig:oneround11}(a), when the base
station can obtain a valid fusion result, the bits transmitted by
the uncompromised nodes to the base station in the one-round scheme
increase with the number of compromised nodes $C$, as expected.
However, they are smaller than those in the witness-based approach.
Notably, in this case, the bits transmitted by uncompromised nodes
in the witness-based approach is constant, since once an
uncompromised node is polled, the polling process is completed. For
small $C$, such as $C=0,1,2$ or $3$, the number of bits transmitted
by uncompromised nodes in the one-round scheme is about half that of
those in the witness-based approach. Additionally, the performance
of the one-round scheme when $P_f=1$ is the worst in all three
simulations, since any compromised node that agrees with the forged
result will sometimes make the forged result with the largest number
of votes and force the base station to use it as the temporary
voting result to poll the next node. Then, the next uncompromised
node needs to transmit its fusion result to the base station instead
of only sending a agreeing vote and the total number of bits
transmitted by the uncompromised nodes increases.

According to Fig.~\ref{fig:oneround11}(b), when the base station
cannot obtain a valid fusion result, the number of bits transmitted
by the uncompromised nodes to the base station in the one-round
scheme decreases with the number of compromised nodes $C$, except
for $P_f=1$. This phenomenon is caused by the fact that the scheme
stops when the number of unpolled nodes plus the maximum number of
votes for any result recorded at the base station is less than $T$.
When $T=M-1$ as simulated, the recoding of two different results at
the base station stops the polling process. Recall that when
$P_f=1$, the only way to stop the polling process is for one fusion
result to be sent by an uncompromised node and the other to be sent
by a compromised node, and for the transmitted bits of the
uncompromised nodes to be the same for all $C$. If $P_f\neq 1$, the
two compromised nodes may yield two different results, and no bit is
transmitted by the uncompromised node. This concludes the simulation
results. This subfigure reveals that the one-round scheme
outperforms when $C$ is small but loses when $C$ exceeds $T/2$.
Similarly, the maximum average number of bits in the one-round
scheme is compared with that of the witness-based approach for
$M=11$ and $12$, and the results are given in
Table~\ref{tbl:max_overhead_or}. This table indicates that the
maximum average number of bits transmitted in the one-round scheme
is much lower than that in the witness-based approach.

\begin{table}
\centering
\renewcommand{\arraystretch}{1.3}
\caption{Maximum average overheads (bits) for one-round (OR) scheme
and witness-based approach given
in~\protect\cite{du:assurance}}\centering
\begin{tabular}{|c||c|c|c|c|}
\hline
$$ & Witness-based & OR, $P_f=0$ & OR, $P_f=0.5$ & OR, $P_f=1$ \\
\hline
$M=11$ & 240 & 71.5 & 86.7 & 128.6 \\
\hline
$M=21$ & 566 & 76.1 & 106.4 & 205.7 \\
\hline
\end{tabular} \label{tbl:max_overhead_or}
\end{table}

\section{Conclusions and Future Work}
\label{sec:conclusions} This work proposes a power-efficient scheme
for data fusion assurance, in which the base station in the wireless
sensor network collects the fusion data and the votes on the data
directly from the fusion nodes. The proposed scheme is more reliable
with less assurance overhead and delay than the witness-based
approach. That is, the power and delay for the transmission of the
fusion result and the votes are significantly decreased.

In the future, we will discover the performance when a node is
compromised with some probability, both statistically dependent and
independent to other nodes. Moreover, the propose scheme cannot be
applied to multi-hop WSNs. We will develop other schemes based on
the direct voting mechanism for the multi-hop WSN.

\appendix

\begin{center}{\bf Performance Analysis of the Variant-round Scheme when
$P_f=1$}\end{center}

In this analysis we assume that $2T+1\ge M$. Since when $C>T$ the
base station will get a forged fusion result and this should be
avoided, we are only considering two cases:
\begin{description}
\item[Case 1]\ $M-T \le C \le T$,
\item[Case 2]\ $C<M-T$.
\end{description}
Case $1$ does not produce a valid fusion result. Assume that the
chosen node at the first round of polling {\bf is} compromised. The
probability that the chosen node is compromised at the first round
is given by $C/M$. The first-round polling-for-vote process finishes
when $M-T$ witness nodes do not agree with the transmitted fusion
result, as described in Step~3. When the polling stops at the $i$th
witness node, the $i$th witness node (uncompromised) does not agree
with the transmitted result and the base station has polled $M-T$
disagreeing nodes (including the $i$th witness node). Moreover,
$M-i-1$ nodes are unpolled, and $M-C-(M-T)=T-C$ of them are
uncompromised. Since the witness set contains $C-1$ compromised
nodes, the maximum number of polled witness nodes is
$(C-1)+(M-T)=M-T+C-1$. Hence, the probability that the polling stops
at the $i$th witness node, where $M-T \le i \le M-T+C-1$, is given
by
\begin{equation*}
P_{v1}^{c1}(i)=\frac{1}{\Pi_{v1}^{c1}}\left(\begin{array}{c} i-1 \\
M-T-1
\end{array}\right) \left(\begin{array}{c} M-i-1 \\ T-C
\end{array}\right).
\end{equation*}
The number of agreeing nodes, $A$, equals to $i-(M-T)$. The first
witness node that disagreed with the transmitted fusion result at
the first round becomes the new chosen node at the second round as
stated in Step~4. The size of the witness set becomes $M^{\prime} =
(M-1)-(i-M+T)-1=2M-T-i-2$. If $M^{\prime}<T$, which is equivalent to
$(M-1)-T-1<A$, then the polling stops as described in the second
part of Step~4. Otherwise, since the previous chosen fusion node is
compromised, the current chosen fusion node is not compromised. At
the second round, the witness set now has
$C^{\prime}=C-1-(i-(M-T))=M+C-T-i-1$ compromised nodes and $M-C-1$
uncompromised nodes. As implied in Step~3, the base station first
polls the $M-T-1$ witness nodes that disagreed with the previous
transmitted fusion result in the first round. The witness set
contains $M-i-1$ unpolled nodes of which $M-C-(M-T)=T-C$ are
uncompromised. Consequently, the number of the possible polling
orders is given by
\[
\Pi_{v1}^{c2} = \left(\begin{array}{c} M-i-1 \\ T-C
\end{array}\right).
\]
When the second round of polling stops at the $j$th witness node,
the $j$th witness node (which is a compromised node) does not agree
with the transmitted result. The base station has polled
$M^{\prime}-T+1$ disagreeing nodes (including the $j$th witness
node) and $j-(M^{\prime}-T+1)=j-M^{\prime}+T-1$ agreeing
(uncompromised) nodes. The first $M-T-1$ of the $j$ polled nodes are
uncompromised, and the rest $j-(M-T-1)-1$ nodes (including the $j$th
witness node) may include the uncompromised or compromised nodes.
Since the $j$ polled nodes include $M^{\prime}-T+1$ disagreeing
(compromised) nodes, the $M^{\prime}-j$ unpolled nodes include
$C^{\prime}-(M^{\prime}-T+1)$ compromised nodes. The probability
that the second polling process stops at the $j$th witness node,
where $j \ge (M^{\prime}-T+1)+(M-T-1)=M^{\prime}+M-2T$ and $j \le
(M^{\prime}-T+1) + (M-C-1) = M-C+ M^{\prime}-T$ (disagreeing nodes
and uncompromised nodes), is given by
\begin{eqnarray*}
P_{v1}^{c2}(j)&=&\frac{1}{\Pi_{v1}^{c2}}\left(\begin{array}{c} j-(M-T-1)-1 \\
M^{\prime}-T
\end{array}\right) \left(\begin{array}{c}  M^{\prime}-j \\
C^{\prime}-(M^{\prime}-T+1)
\end{array}\right) \nonumber \\
&=&\frac{1}{\Pi_{v1}^{c2}}\left(\begin{array}{c} j-M+T \\ 2M-2T-i-2
\end{array}\right) \left(\begin{array}{c}  2M-T-i-j-2 \\ T+C-M
\end{array}\right).
\end{eqnarray*}
Since $A$ is at least $M-T-1$ at the second round, it is easy to see
that $M^{\prime}-T-1<A$ and the scheme will stop after the second
round. Therefore, the average overhead is
\begin{equation}
O_{v1}^c(M,T,C,1) = \sum_{i=M-T}^{M-T+C-1} P_{v1}^{c1}(i)
\left[(M-T)k^{\prime} + \sum_{j=M^{\prime}+M-2T}^{M-C+
M^{\prime}-T}P_{v1}^{c2}(j) (j-M^{\prime}+T-1)k \right].
\label{eqn:polling_invalid_compromised_overhead}
\end{equation}

As mentioned before, when $M^{\prime}<T$, i.e., $2M-2T-2<i$, the
scheme will stop after the first round. Hence, the average round
delay is
\begin{equation}
R_{v1}^c(M,T,C,1) = \sum_{i=M-T}^{2M-2T-2} 2P_{v1}^{c1}(i)
 + \sum_{i=2M-2T-1}^{M-T+C-1}P_{v1}^{c1}(i),
\label{eqn:polling_invalid_compromised_round_delay1}
\end{equation}
and the average polling delay is
\begin{equation}
D_{v1}^c(M,T,C,1) = \sum_{i=M-T}^{2M-2T-2}P_{v1}^{c1}(i)
\left[i+\sum_{j=M^{\prime}+M-2T}^{M-C+ M^{\prime}-T}jP_{v1}^{c2}(j)
\right]
 + \sum_{i=2M-2T-1}^{M-T+C-1}iP_{v1}^{c1}(i).
\label{eqn:polling_invalid_compromised_polling_delay1}
\end{equation}

 In the first case, the probability
that the chosen node {\bf is not} compromised at the first round is
given by $(M-C)/M$. The number of the possible polling orders is
written as
\[
\Pi_{v1}^{u1} = \left(\begin{array}{c} M-1 \\ C \end{array}\right).
\]
Thus, the probability that the polling stops at the $i$th witness
node, where $M-T \le i \le (M-C-1)+(M-T)=2M-T-C-1$, is expressed as
\begin{equation*}
P_{v1}^{u1}(i)=\frac{1}{\Pi_{v1}^{u1}}\left(\begin{array}{c} i-1 \\
M-T-1
\end{array}\right) \left(\begin{array}{c} M-i-1 \\ T+C-M
\end{array}\right).
\end{equation*}
At the second round the base station chooses the first witness nodes
that disagreed with the transmitted fusion result sent in the
previous round. The size of the witness set becomes
$M^{\prime}=2M-T-i-2$. The witness set has $M-C-(i-(M-T))-1 =
2M-C-T-i-1$ uncompromised nodes. Similarly, the polling stops if
$M^{\prime}<T$. Otherwise, since the previous chosen fusion node is
not compromised, the current chosen fusion node is compromised.
Thus, the number of the possible polling orders is given by
\[
\Pi_{v1}^{u2} = \left(\begin{array}{c} M-i-1 \\ T+C-M
\end{array}\right).
\]
When the second round of polling stops at the $j$th witness node,
the $j$ polled nodes contain $M^{\prime}-T+1$ disagreeing
(uncompromised) nodes and the unpolled nodes include
$2M-C-T-i-1-(M^{\prime}-T+1)$ uncompromised nodes. The probability
that the second round of polling finishes at the $j$th witness node,
where $(M^{\prime}-T+1)+(C-1)=M^{\prime}-T+C \ge j \ge
(M^{\prime}-T+1)+(M-T-1)=M^{\prime}+M-2T$, is given by
\begin{eqnarray*}
P_{v1}^{u2}(j)&=&\frac{1}{\Pi_{v1}^{u2}}\left(\begin{array}{c} j-(M-T-1)-1 \\
M^{\prime}-T
\end{array}\right) \left(\begin{array}{c}  M^{\prime}-j \\
2M-C-T-i-1-(M^{\prime}-T+1)
\end{array}\right) \nonumber \\
&=&\frac{1}{\Pi_{v1}^{u2}}\left(\begin{array}{c} j-M+T \\ 2M-2T-i-2
\end{array}\right) \left(\begin{array}{c}  2M-T-i-j-2 \\ T-C
\end{array}\right).
\end{eqnarray*}
Similarly, it is easy to see that the scheme will stop after the
second round. Therefore, the average overhead is given by
\begin{equation}
O_{v1}^{u}(M,T,C,1) = \sum_{i=M-T}^{2M-T-C-1} P_{v1}^{u1}(i)
\left[(i-M+T)k+ \sum_{j=M^{\prime}+M-2T}^{M^{\prime}-T+C}
P_{v1}^{u2}(j)(M^{\prime}-T+1)k^{\prime} \right].
 \label{eqn:polling_invalid_noncompromised_overhead}
\end{equation}
Again, when $M^{\prime}<T$, i.e., $2M-2T-2<i$, the scheme will stop
after the first round. Hence, the average round delay is
\begin{equation}
R_{v1}^u(M,T,C,1) = \sum_{i=M-T}^{2M-2T-2} 2P_{v1}^{u1}(i)
 + \sum_{i=2M-2T-1}^{2M-T-C-1}P_{v1}^{u1}(i),
\label{eqn:polling_invalid_uncompromised_round_delay1}
\end{equation}
and the average polling delay is
\begin{equation}
D_{v1}^u(M,T,C,1) = \sum_{i=M-T}^{2M-2T-2}P_{v1}^{u1}(i)
\left[i+\sum_{j=M^{\prime}+M-2T}^{M^{\prime}-T+C}jP_{v1}^{u2}(j)
\right]
 + \sum_{i=2M-2T-1}^{M-T+C-1}iP_{v1}^{u1}(i).
\label{eqn:polling_invalid_uncompromised_polling_delay1}
\end{equation}

From (\ref{eqn:polling_invalid_compromised_overhead}) and
(\ref{eqn:polling_invalid_uncompromised_polling_delay1}), we have
the average overhead and the average delays of Case 1 as

\begin{equation}
O_{v1}(M,T,C,1) = \frac{C}{M} O_{v1}^c(M,T,C,1) + \frac{M-C}{M}
O_{v1}^u(M,T,C,1), \label{eqn:polling_invalid_overhead}
\end{equation}
\begin{equation}
R_{v1}(M,T,C,1) = \frac{C}{M} R_{v1}^c(M,T,C,1) + \frac{M-C}{M}
R_{v1}^u(M,T,C,1), \label{eqn:polling_invalid_round_delay1}
\end{equation}
and
\begin{equation}
D_{v1}(M,T,C,1) = \frac{C}{M} D_{v1}^c(M,T,C,1) + \frac{M-C}{M}
D_{v1}^u(M,T,C,1). \label{eqn:polling_invalid_polling_delay1}
\end{equation}

A valid fusion result is available in the second case. The
first-round process is similar to the first-round process in the
first case when the chosen node {\bf is} compromised at the first
round. The probability that the $T$th witness node agrees with the
transmitted fusion result at witness node $j$ of the second-round
polling process is
\begin{eqnarray*}
P_{v2}^{c2}(j)&=&\frac{1}{\Pi_{v1}^{c2}}\left(\begin{array}{c} j-(M-T-1)-1 \\
T-(M-T-1)-1
\end{array}\right) \left(\begin{array}{c}  2M-T-i-j-2 \\
M-C-(T+1)
\end{array}\right) \nonumber \\
&=&\frac{1}{\Pi_{v1}^{c2}}\left(\begin{array}{c} j-M+T \\ 2T-M
\end{array}\right) \left(\begin{array}{c}  2M-T-i-j-2 \\ M-C-T-1
\end{array}\right).
\end{eqnarray*}
Significantly, if $M=2T+1$, then $j=T$. Otherwise, $T \le j \le
C-(i-(M-T))-1+T=M+C-i-1$. Therefore, the average overhead is
expressed as
\begin{eqnarray}
O_{v2}^c(M,T,C,1) = \left\{ \begin{array}{ll} \sum_{i=M-T}^{M-T+C-1}
P_{v1}^{c1}(i) \left[(M-T)k^{\prime} + Tk \right]=(M-T)k^{\prime} + Tk& M=2T+1 \\
\sum_{i=M-T}^{M-T+C-1} P_{v1}^{c1}(i) \left[(M-T)k^{\prime} +
\sum_{j=T}^{M+C-i-1}P_{v2}^{c2}(j) Tk \right]& \mbox{else}
\end{array} \right..
 \label{eqn:polling_valid_compromised_overhead}
\end{eqnarray}
Hence, the average round delay is
\begin{equation}
R_{v2}^c(M,T,C,1) = \sum_{i=M-T}^{M-T+C-1} 2P_{v1}^{c1}(i)
\label{eqn:polling_valid_compromised_round_delay1}
\end{equation}
and the average polling delay is
\begin{equation}
D_{v2}^c(M,T,C,1) = \sum_{i=M-T}^{M-T+C-1}P_{v1}^{c1}(i)
\left[i+\sum_{j=T}^{M+C-i-1}jP_{v2}^{c2}(j) \right].
\label{eqn:polling_valid_compromised_polling_delay1}
\end{equation}
 Only one round of polling is needed when
the chosen node is uncompromised at the first round. The probability
that the polling process ends at the witness node $i$, where $T\le i
\le T+C$, is written as
\begin{equation*}
P_{v2}^{u1}(i) = \frac{1}{\Pi_{v1}^{u1}}\left(\begin{array}{c} i-1 \\
T-1
\end{array}\right) \left(\begin{array}{c} M-i-1 \\ M-C-T-1
\end{array}\right).
\end{equation*}
The average overhead, and the delays are given by
\begin{equation}
O_{v2}^u(M,T,C,1)=\sum_{i=T}^{T+C} P_{v2}^{u1}(i) Tk,
\label{eqn:polling_valid_noncompromised_overhead}
\end{equation}
\begin{equation}
R_{v2}^u(M,T,C,1)=1,
\label{eqn:polling_valid_noncompromised_round_delay1}
\end{equation}
and
\begin{equation}
D_{v2}^u(M,T,C,1)=\sum_{i=T}^{T+C} iP_{v2}^{u1}(i).
\label{eqn:polling_valid_noncompromised_polling_delay1}
\end{equation}

{From} (\ref{eqn:polling_valid_compromised_overhead}) and
(\ref{eqn:polling_valid_noncompromised_polling_delay1}), the average
overhead, the delays in Case 2 are given by
\begin{eqnarray*}
O_{v2}(M,T,C,1) = \frac{C}{M} O_{v2}^c(M,T,C,1) + \frac{M-C}{M}
O_{v2}^u(M,T,C,1),
\end{eqnarray*}
\begin{eqnarray*}
R_{v2}(M,T,C,1) = \frac{C}{M} R_{v2}^c(M,T,C,1) + \frac{M-C}{M}
R_{v2}^u(M,T,C,1),
\end{eqnarray*}
and
\begin{eqnarray*}
D_{v2}(M,T,C,1) = \frac{C}{M} D_{v2}^c(M,T,C,1) + \frac{M-C}{M}
D_{v2}^u(M,T,C,1).
\end{eqnarray*}

\end{document}